\documentclass[12pt]{article}

\usepackage{epsfig}
\usepackage{psfrag}

\def\ZZ{{\mathchoice {\hbox{$\sf\textstyle Z\kern-0.4em Z$}}
{\hbox{$\sf\textstyle Z\kern-0.4em Z$}}
{\hbox{$\sf\scriptstyle Z\kern-0.3em Z$}}
{\hbox{$\sf\scriptscriptstyle Z\kern-0.2em Z$}}}}

\makeatletter
\long\def\@makecaption#1#2{
    \vskip 20pt
    \setbox\@tempboxa\hbox{\footnotesize {\bf #1} #2}
    \dimen0=\hsize
    \advance\dimen0 by -10pt
    \ifdim \wd\@tempboxa >\dimen0
        \hbox to \hsize{
            \hfil
            \parbox{12cm}{\def\baselinestretch{1.05}\footnotesize
                    {\bf #1} #2
                    }
                    \hfil}
    \else \hbox to \hsize{\hfil \box\@tempboxa \hfil}
    \fi}
\makeatother

\begin{document}
\begin{titlepage}
\begin{flushright}
  FAU-TP3-00/2\\
hep-th/0003195    
\end{flushright}
\vskip 3.0cm
\begin{center}
{\Large {\bf Emergence of Skyrme crystal in Gross-Neveu
\vskip 0.1cm
and 't~Hooft models at finite density}}
\vskip 0.5cm
Verena Sch\"on and Michael Thies
\vskip 0.2cm
{\it Institute for Theoretical Physics III,
University of Erlangen-N\"urnberg, \\
 Staudtstr. 7, 91058 Erlangen, Germany}
\end{center}
\vskip 3.0cm
\begin{abstract}
We study two-dimensional, large $N$ field theoretic models (Gross-Neveu
model, 't~Hooft model) at finite baryon density near the chiral
limit. The same mechanism
which leads to massless baryons in these models induces a breakdown
of translational invariance at any finite density. In the chiral limit
baryonic matter
is characterized by a spatially varying chiral angle with a wave number
depending only on the density.
For small bare quark masses
a sine-Gordon kink chain is obtained which may be regarded
as simplest realization of the Skyrme crystal for nuclear
matter. Characteristic differences between
confining and non-confining models are pointed out.
\end{abstract}

\end{titlepage}
\sloppy

\section*{1) Introduction}

The description of baryonic matter on the basis of QCD remains a
theoretical
challenge, especially since lattice gauge calculations have
so far been of little help for this problem.
Nuclear physics, relativistic heavy-ion physics and astrophysics are
some of the fields which would greatly benefit
from any progress on this issue. Recently, a new surge of
interest has been triggered by the suggestion that at high density,
the novel phenomenon of color superconductivity might set in
\cite{Alford,Rapp}. This development has highlighted how little is
known reliably about dense, strongly interacting matter.

Here, we address a much simpler finite density
problem where a full analytic solution
can be found: We consider two-dimensional model field theories with
interacting fermions at or near the chiral limit. Specifically, we
have in mind
the two-dimensional version of the Nambu--Jona-Lasinio model \cite{Nambu},
{\em i.e.}, the chiral Gross-Neveu model \cite{Gross}, and QCD$_2$ with
fundamental quarks, the 't~Hooft model \cite{tHooft74b}.
In both cases, one considers a large number $N$ of fermion species
and investigates the limit $N\to \infty$, keeping $Ng^2$ fixed
\cite{tHooft74a}.
These two models are quite similar as far as their chiral properties are
concerned but differ with respect to confinement of quarks which is
only exhibited by the 't~Hooft model. So far, the phase diagram
of the Gross-Neveu model has been studied extensively as function
of temperature and chemical potential
\cite{Harrington,Dashen,Wolff,Treml,Barducci},
and the results seem to be uncontroversial. The 't~Hooft model
on the other hand has been investigated only sporadically at finite
temperature \cite{McLerran,Li86}, most recently in Ref.~\cite{Schoen},
but hardly anything is known yet about its properties at
finite density \cite{Li86}.

Our point of departure is the following observation: Both of these
models possess light baryons whose mass vanishes in the chiral
limit \cite{Affleck,Salcedo,Lenz} (by light, we mean that $M_B/N$ is
small on the relevant physical scale). This is of course no accident,
but a generic feature of models with broken chiral symmetry in 1+1
dimension --- the baryons are topologically non-trivial excitations
of the Goldstone boson (``pion") field.
By contrast, the Gross-Neveu model with discrete chiral symmetry can only
accommodate baryons whose mass scales with the physical fermion
mass and hence stays finite in the chiral limit. Nevertheless, it has
been argued that both variants of the Gross-Neveu model have identical
phase diagrams \cite{Barducci}.
Since we find it rather perplexing that the structure of
the single baryon should have no influence on the structure of baryonic
matter, we have reinvestigated this issue. We have found that a
combination of large $N$ techniques with the strong constraints arising from
broken chiral symmetry is powerful
enough to allow for a simple, analytic solution of this problem
in the vicinity of the chiral limit. The results of our analysis differ
qualitatively from the conventional wisdom about the Gross-Neveu model
and carry over to the (confining) 't~Hooft model as well.
They seem to confirm a number of investigations of other field theories
at finite fermion density, where strikingly similar behaviour was
found. These include exact studies of two-dimensional models
like the massive \cite{Fischler79} and massless Schwinger
model \cite{Kao94} or finite $N_c$ and
$N_f$ massless QCD$_2$ \cite{Christiansen96} as well as more
approximate treatments of
4-dimensional large $N$ QCD \cite{Deryagin92,Shuster99,Park99} and
effective chiral quark models \cite{Kutschera90,Sadzikowski00}.

A word of caution is in order here: Throughout this paper,
we shall constantly deal with spontaneous
breaking of continuous symmetries and Goldstone bosons in 1+1 dimensions,
in seeming conflict with the Coleman-Mermin-Wagner theorem
\cite{Coleman73,Mermin}. It is well
understood by now that the large $N$ limit enables one to circumvent this
no-go theorem. As clearly explained by Witten \cite{Witten78},
the bad infrared behaviour of the boson propagator, when exponentiated,
gives rise to a power law correlator $|x-y|^{-1/N}$ which becomes constant
in the limit $N\to \infty$. This kind of almost long-range order is also
familiar from the two dimensional $XY$-model \cite{Berezinski,Kosterlitz}.
Alternatively, one may argue that the
mean field approximation predicts symmetry breakdown and that this result is
protected against fluctuations (which would otherwise restore the
symmetry in two dimensions) by $1/N$ suppression factors.
In this sense, low dimensional large $N$ theories are not only more
tractable, but also physically more appealing than their finite $N$
counterparts. They bear more resemblance to the real, 3+1 dimensional world.

This paper is organized as follows. In Sect.~2 we briefly review the
conventional analytical treatment of the chiral Gross-Neveu model at finite
density  and
point out a certain deficiency of this approach. In Sect.~3 we repeat a similar
analysis for the 't~Hooft model, supplementing the analytical
methods by numerical computations where necessary. In Sect.~4 the Skyrme
\cite{Skyrme} type
of approach to the light baryons in both models is recalled
\cite{Salcedo} and
generalized to the case of baryonic matter in the strict chiral limit.
In Sect.~5 we then allow for a small symmetry breaking mass term and
make contact with the sine-Gordon kink chain, the two dimensional
analog of the Skyrme crystal \cite{Klebanov}. This is followed by a short
summary and conclusions in Sect.~6.

\section*{2) Chiral Gross-Neveu model at finite density: Conventional
approach}

Let us first recall the standard treatment of the Gross-Neveu model at
finite density. In the large $N$ limit mean-field techniques become
exact. Technically, they may be phrased in a variety of ways.
We choose the language of relativistic many-body theory, following
Refs.~\cite{Salcedo,Lenz,Pausch}, which we find particularly
intuitive for the
problem at hand. Then the vacuum,
the baryon and baryonic matter are all described by a relativistic
Hartree-Fock approach (for baryons in the large $N$ limit this was
first recognized in Ref.~\cite{Witten79}). ``Conventional approach"
in the title of this section refers to translational invariance ---
we shall assume that the system is described by an interacting Fermi gas
with prescribed, homogeneous density. We shall first deal with
the Gross-Neveu model with continuous
chiral symmetry ($\psi \to
{\rm e}^{{\rm i}\alpha \gamma_5}\psi$) and Lagrangian density \cite{Gross}
\begin{equation}
{\cal L} = \bar{\psi} {\rm i} \partial \!\!\!/ \psi + \frac{1}{2}g^2
\left( (\bar{\psi}\psi)^2+(\bar{\psi}{\rm i}\gamma_5 \psi)^2 \right) \ .
\label{d0}
\end{equation}
As a matter of fact, the corresponding calculation would
be identical for the
model with discrete chiral symmetry only ($\psi\to \gamma_5 \psi$), where
the $\gamma_5$-term in Eq.~(\ref{d0}) is omitted. The results
presented here are well known, but our aim is to criticize them
in a novel way.

We denote the fermion density per color (or baryon density) by $\rho_B
= p_f/\pi$
($p_f$: Fermi momentum). At the mean field level, the fermions acquire
a physical mass $m$ which has to be determined self-consistently.
The ground state
energy density per color is given by
\begin{equation}
\frac{{\cal E}}{N}=-2 \int_{p_f}^{\Lambda/2} \frac{{\rm d}k}{2\pi} \sqrt{
m^2+k^2} + \frac{m^2}{2N g^2}
\label{d1}
\end{equation}
where $\Lambda$ is an ultra-violet cutoff.
The first term is just the sum over single particle energies for
all occupied states (the Dirac sea plus all positive energy states
with $|p|<p_f$), the second term the usual correction for double
counting of interaction effects familiar from the Hartree-Fock approximation.
To renormalize the theory, let us
first consider the limit $p_f\to 0$, denoting the physical fermion mass
in the vacuum by $m_0$,
\begin{equation}
\frac{{\cal E}}{N}=-2 \int_{0}^{\Lambda/2} \frac{{\rm d}k}{2\pi} \sqrt{
m_0^2+k^2} + \frac{m_0^2}{2N g^2} \ .
\label{d2}
\end{equation}
Minimizing ${\cal E}$ with respect to $m_0$ yields the relativistic
Hartree-Fock equation
\begin{equation}
m_0\left(1+\frac{Ng^2}{\pi}\ln \frac{m_0}{\Lambda} \right) = 0 \ .
\label{d3}
\end{equation}
Due to the similarity in structure between the relativistic Hartree-Fock
approach
and BCS theory \cite{Bardeen}, this is often referred to as the
``gap equation".
The non-trivial solution (which has always
lower vacuum energy) yields the relation
\begin{equation}
\frac{Ng^2}{\pi}\ln \frac{\Lambda}{m_0} = 1
\label{d4}
\end{equation}
which teaches us how the bare coupling constant depends on the
cutoff parameter, given $m_0$. Recall that the Gross-Neveu model shares
with real QCD both asymptotic freedom and dimensional transmutation; these
properties are contained in Eq.~(\ref{d4}).
Using this relation to renormalize the matter ground state energy density,
Eq.~(\ref{d1}), we find (dropping an irrelevant term $-\Lambda^2/8 \pi$)
\begin{equation}
\frac{{\cal E}}{N}= -\frac{m^2}{4\pi}
+ \frac{1}{2\pi} p_f \sqrt{p_f^2+m^2}
+ \frac{1}{2\pi} m^2 \ln \left( \frac{p_f + \sqrt{m^2+p_f^2}}{m_0}\right) \ .
\label{d5}
\end{equation}
The energy is minimal provided $m$ satisfies
\begin{equation}
m \ln \left( \frac{p_f + \sqrt{m^2+p_f^2}}{m_0}\right)=0 \ ,
\label{d6}
\end{equation}
{\em i.e.}, for
\begin{equation}
m=0 \ \quad \mbox{or} \quad m= m_0\sqrt{1-
\frac{2 p_f}{m_0}} \quad \left(p_f< \frac{m_0}{2}\right) \ .
\label{d7}
\end{equation}
The corresponding energy densities are
\begin{eqnarray}
\left. \frac{{\cal E}}{N}\right|_{m=0} &=& \frac{p_f^2}{2\pi} \ ,
\nonumber \\
\left. \frac{{\cal E}}{N} \right|_{m\neq 0}& = & -\frac{m_0^2}{4\pi} + \frac{p_f m_0}
{\pi} -\frac{p_f^2}{2\pi} \qquad \left(p_f < \frac{m_0}{2}\right)\ .
\label{d8}
\end{eqnarray}
The physical quark masses (\ref{d7}) and the energy densities (\ref{d8}) are
plotted in Figs. 1 and 2. From these figures one might be tempted to
conclude that chiral symmetry is broken at low densities and
gets restored in a second order
phase transition at $p_f=m_0/2$. As is well known,
this does not occur, rather there is a first order chiral phase transition
at $p_f=m_0/\sqrt{2}$. This can easily be inferred by inspection of the
thermodynamic potential of the Gross-Neveu model \cite{Wolff}. For our
purpose, the following physical reasoning is perhaps more instructive:
Let us compare the energy densities (\ref{d8}) with the energy
density for a system of size $L$ divided into two homogeneous regions I
(size $\ell$) and II (size $L-\ell$). In region I chiral symmetry is
restored; it contains the extra fermions needed to get the prescribed
average density (the ``MIT bag" \cite{Chodos}). Region II
consists of the physical vacuum
with broken chiral symmetry, void of excess fermions. The mean energy
density obtained in this way is
\begin{equation}
\frac{{\cal E}}{N} = -\left(\frac{L-\ell}{L}\right) \frac{m_0^2}{4\pi}
+ \frac{L p_f^2}
{2\pi \ell} \ .
\label{d9}
\end{equation}
Minimization with respect to $\ell$ yields
\begin{equation}
\ell = \frac{\sqrt{2} p_f L}{m_0}
\label{d10}
\end{equation}
valid for $p_f<m_0/\sqrt{2}$, and hence the optimal energy density
\begin{equation}
\frac{{\cal E}}{N} = - \frac{m_0^2}{4\pi} + \frac{p_f m_0}{\sqrt{2}\pi} \qquad
\left(p_f < \frac{m_0}{\sqrt{2}} \right) \ .
\label{d11}
\end{equation}
As shown in Fig.~2, this solution is lower in energy than the homogeneous one;
moreover, it yields the convex hull of ${\cal E}$.
It ends exactly at the first order phase transition point $p_f=
m_0/\sqrt{2}$ where all space is filled with one big bag.
This should be contrasted to the scenario underlying Fig.~1 where the
fermion mass decreases continuously. We thus recover
the generally accepted mixed phase interpretation of the
Gross-Neveu model at finite density.
Notice also that only the total size of regions I and II matters, not how they
are subdivided; there could be baryon ``droplets" as well. Alternatively,
the mixed phase curve in Fig.~2 with its linear dependence on $p_f$ could
have been inferred from a standard Maxwell construction.
It is interesting that
a very similar qualitative behaviour was found recently in 3+1 dimensions,
where the close relationship with the bag model was also
stressed \cite{Alford}.

One important point to which we would like to draw the attention
of the reader is
the behaviour of ${\cal E}$ near $\rho_B=p_f/\pi=0$. Since ultimately, at
very low density, the fermionic matter problem must reduce
to the problem of a single baryon, one would expect
\begin{equation}
\left. \frac{\partial {\cal E}}{\partial \rho_B}\right|_{\rho_B=0} =
M_B
\label{d12}
\end{equation}
where $M_B$ is the baryon mass.
In the present calculation,
$M_B$ is not the physical baryon mass, but the mass of an alleged
``delocalized" baryon. This is inherent in the translationally invariant
Hartree-Fock approach, {\em i.e.}, the
assumption that the single particle orbitals are momentum eigenstates.
Using Eq.~(\ref{d12}) we obtain in
the homogeneous, single phase calculation, Eq.~(\ref{d8}),
$M_B=Nm_0$, consistent with a short range force and a delocalized baryon.
The (physically more viable) mixed phase approach, Eq.~(\ref{d11}),
predicts a baryon mass lower by a factor of $1/\sqrt{2}$.
This factor can readily be understood in terms of the bag model. Indeed,
it follows from Eq.~(\ref{d9}) that ($E={\cal E}L$)
\begin{equation}
E_B-E_0 = N \ell  \left( \frac{m_0^2}{4\pi} + \frac{q_f^2}{2\pi} \right) \ , \quad
q_f= \frac{\pi B}{\ell} \ .
\label{d13}
\end{equation}
For $B=1$,
this expression can be interpreted as energy of a single
baryon, $\ell$ being its diameter.
The first term is just the bag pressure (the difference between the energy
density
of the physical vacuum and that of the perturbative one, cf. Eq.~(\ref{d8})
for $p_f=0$), the
second the kinetic energy of $N$ massless
quarks. The bag size $\ell$ is found through minimization of
the energy (for $B=1$) to be
\begin{equation}
\ell = \frac{\sqrt{2}\pi}{m_0}.
\label{d14}
\end{equation}
Inserting this result into Eq.~(\ref{d13}), one finds that the bag pressure and
the quark kinetic energy contributions
are exactly equal in this model and that $M_B=N m_0/\sqrt{2}$.

However, the Gross-Neveu model possesses bound baryons with lowest mass
$Nm_0/\pi$ (kink solution for the model with discrete chiral symmetry
\cite{Gross,Pausch}), or even massless baryons (model with continuous chiral
symmetry \cite{Salcedo}). These binding effects are not $1/N$ suppressed
and should be
correctly reproduced in a Hartree-Fock approach, in the low density limit.
They have obviously been missed here due to our tacit assumption of
translational invariance.
There is no good reason why such effects should not play a role at higher
densities as well. Moreover,
differences between the continuous and discrete chirally
symmetric Gross-Neveu models based on their different baryon structure
and masses are not at all captured by the ``conventional" approach.
Below, we shall present a
cure for this disease. Before that however,
let us first repeat the naive calculation for the 't~Hooft model,
where the corresponding results are not yet available in the
literature.

\section*{3) 't~Hooft model at finite density, assuming
translational invariance}

The 't~Hooft model is defined as the large $N$ limit of 1+1 dimensional
SU($N$) gauge theory with quarks in the fundamental representation
\cite{tHooft74b},
\begin{equation}
{\cal L} = \bar{\psi} {\rm i} D \!\!\!\!/ \, \psi - \frac{1}{2}{\rm tr}
F_{\mu \nu} F^{\mu \nu} \ .
\label{d14a}
\end{equation}
Since the
light-cone approach originally used by 't~Hooft to determine
the meson spectrum seems to be less convenient for the vacuum, baryon
and baryonic matter problems, we shall work in normal coordinates.
This approach was pioneered by Bars and Green \cite{Bars} and further
developed in Refs.~\cite{Li87,Salcedo,Lenz}. Common to all of these works
is the fact that the gluons are gauged away (axial gauge), leaving
behind a theory of fermions interacting via a linear Coulomb
potential. We refer the reader to the detailed derivation of the
Hartree-Fock approach in Refs.~\cite{Salcedo,Lenz} and immediately
proceed to the formulae which are relevant for our purpose. Let us first
summarize the treatment of the vacuum. A central quantity
is the single particle density matrix in momentum space,
\begin{equation}
\rho(p) = \frac{1}{2} + \gamma^0 \rho_0(p)-{\rm i}\gamma^1 \rho_1(p)
+\gamma^5 \rho_5(p) \ .
\label{d15}
\end{equation}
Its precise definition in terms of the quark fields is
\begin{equation}
\rho_{\alpha \beta}(p) = \int {\rm d}x\, {\rm e}^{-{\rm i} px}
\langle 0 | \frac{1}{N} \sum_i \psi_{i\beta}^{\dagger}(0)\psi_{i \alpha}
(x) |0\rangle \ .
\label{d15a}
\end{equation}
The Slater determinant condition characteristic for Hartree-Fock,
\begin{equation}
\rho^2(p)=\rho(p) \ \rightarrow  \ \rho_0^2(p)+\rho_1^2(p)+\rho_5^2(p)=1/4
\label{d16}
\end{equation}
holds manifestly in the parametrization
\begin{equation}
\left( \begin{array}{c}\rho_0(p) \\ \rho_1(p)  \\ \rho_5(p) \end{array}
\right) = -\frac{1}{2} \left( \begin{array}{c} \sin \theta(p) \cos \varphi
\\ \sin \theta(p) \sin  \varphi \\ \cos \theta(p) \end{array} \right)\ .
\label{d17}
\end{equation}
$\theta(p)$ is the Bogoliubov angle, $\varphi$ the global angle which
locates the
broken symmetry vacuum on the chiral circle (hence it has no $p$-dependence).
The Bogoliubov angles are the variational parameters; this terminology
stems once more from BCS theory, which has the same formal
structure as relativistic
Hartree-Fock.
We choose $\varphi=0$, the value
reached if one lets the bare quark mass approach zero starting from a finite
value. Then,
\begin{equation}
\rho(p) = v(p) v^{\dagger}(p)
\label{d18}
\end{equation}
with the (positive and negative energy) Hartree-Fock spinors
\begin{equation}
u(p)=\left(\begin{array}{r} \cos \theta(p)/2 \\ \sin \theta(p)/2
\end{array} \right) \ , \quad
v(p)=\left(\begin{array}{r} -\sin \theta(p)/2 \\ \cos \theta(p)/2
\end{array} \right) \ .
\label{d19}
\end{equation}
The vacuum expectation value of the Hamiltonian density reads
\begin{equation}
\frac{{\cal E}}{N}=-\int \frac{{\rm d}p}{2\pi} p \cos \theta (p) -
\frac{Ng^2}{8} \int \frac{{\rm d}p}{2\pi}\int \frac{{\rm d}p'}{2\pi}\frac{
\cos(\theta(p)-\theta(p'))-1}{(p-p')^2}
\label{d20}
\end{equation}
where the first term is the kinetic energy, the second the Coulomb
interaction of the quarks.
Varying with respect to the Bogoliubov angles $\theta(p)$, the
gap equation is obtained in the form
\begin{equation}
p \sin \theta(p) + \frac{Ng^2}{4} \int \!\!\!\!\!\!- \frac{{\rm d}p'}{2\pi}
\frac{\sin \left(\theta(p)-\theta(p')\right)}{(p-p')^2} = 0 \ .
\label{d21}
\end{equation}
The integral has to be defined as principal value integral, cf. Refs.~{\cite{Bars,
Li87,Lenz}. We shall also need the expression for
the quark condensate in the vacuum,
\begin{equation}
\langle \bar{\psi}\psi \rangle_{\rm v} = -N \int \frac{{\rm d}p}{2\pi}
\sin \theta(p)\ ,
\label{d22}
\end{equation}
and the quark single particle energies,
\begin{equation}
\omega(p) = p \cos \theta(p) + \frac{Ng^2}{4} \int \frac{{\rm d}p'}
{2\pi} \frac{\cos \left( \theta(p)-\theta(p') \right)}{(p-p')^2}\ .
\label{d23}
\end{equation}
Although the gap equation (\ref{d21}) for the 't~Hooft model must be
solved numerically, the value of the quark condensate (\ref{d22})
is known analytically, owing to an indirect determination via sum rules
and the 't~Hooft equation for mesons \cite{Zhitnitsky};
it is
\begin{equation}
\langle \bar{\psi} \psi \rangle_{\rm v} = - \frac{N}{\sqrt{12}}
\left( \frac{Ng^2}{2\pi} \right)^{1/2} \ .
\label{d23a}
\end{equation}
The single particle energies (\ref{d23}) are badly infrared divergent,
a source of a long and ongoing debate in the literature \cite{Abdalla}.
To exhibit the divergence, we follow Ref.~\cite{Lenz}, isolate the divergent
part of the integral and regularize it by using a finite box of length $L$,
\begin{equation}
\omega(p) = p \cos \theta(p) + \frac{Ng^2}{4} \int \frac{{\rm d}p'}
{2\pi}\frac{\cos \left( \theta(p)-\theta(p') \right)-1}
{(p-p')^2}
+ \frac{Ng^2L}{48} \ .
\label{d24}
\end{equation}
This last constant diverges for $L\to \infty$ but seems to be essential
to account for confinement in such an independent particle
picture: The isolated quarks behave roughly as if they had infinite mass.
If one simply
throws the infinite constant away, as is often done, one gets an awkward
sign change in
$\omega(p)$ at some low momentum $p$, cf. Refs.~\cite{Li87,Salcedo,Lenz},
and runs into serious inconsistencies in finite
temperature Hartree-Fock calculations
\cite{Li86,Schoen}. Fortunately, the constant drops out of the
calculation of color singlet mesons, as already noticed by 't~Hooft (his
IR cutoff parameter $\lambda$ is related to our box size $L$ by
$\lambda=12/\pi L$, cf. Ref.~\cite{Lenz}): The infinite self-energy term
is cancelled by an equally infinite piece in the Coulomb interaction.
We also note in passing that if one employs a finite box as
infrared regulator, one is unambiguously led to 't~Hooft's treatment
of the quark self-energies rather than to Wu's alternative regularization
prescription \cite{Wu,Abdalla}.
Since the emergence of the constant $Ng^2 L/48$ in the single particle
energy (\ref{d24}), but not in the vacuum energy (\ref{d20}), is rather
important for our discussion and somewhat hidden in Ref.~\cite{Lenz},
we have included a simplified version of the arguments underlying
Eqs.~(\ref{d20}) and (\ref{d24}) in the appendix.

After this review of the treatment of the vacuum, we are in a position
to include a finite baryon density, assuming translational invariance.
If $p_f$ denotes the Fermi momentum, we have to replace the density matrix
(\ref{d18}) by
\begin{eqnarray}
\rho(p)&=&\Theta(p_f-|p|)u(p)u^{\dagger}(p) + v(p)v^{\dagger}(p)
\nonumber \\
& = & \Theta(p_f-|p|) + \Theta(|p|-p_f) v(p)v^{\dagger}(p)
\label{d25}
\end{eqnarray}
where we have used the completeness relation for the spinors in
the second step. In the expression for the Hartree-Fock
ground state energy density (\ref{d20}), according to the second line of
Eq.~(\ref{d25}), we must exclude the region
$[-p_f,p_f]$ from the momentum integrations and pick up an additional
term due to the change in the baryon density ${\rm tr} \rho$,
\begin{eqnarray}
\frac{{\cal E}}{N} &=& - \int \frac{{\rm d}p}{2\pi} \Theta(|p|-p_f) p \cos \theta(p)
\nonumber \\
& & - \frac{Ng^2}{8} \int \frac{{\rm d}p}{2\pi} \int \frac{{\rm d}p'}{2\pi}
\Theta(|p|-p_f)
\Theta(|p'|-p_f) \frac{\cos (\theta(p)-\theta(p'))
-1}{(p-p')^2}
\nonumber \\
& & +\frac{Ng^2}{4}\int \frac{{\rm d}p}{2\pi}\int \frac{{\rm d}p'}{2\pi}
\Theta(p_f-|p|)\Theta(|p'|-p_f) \frac{1}{(p-p')^2} \ .
\label{d26}
\end{eqnarray}
This yields at once the following finite density generalization
of the gap equation,
\begin{equation}
p \sin \theta(p) + \frac{Ng^2}{4} \int \!\!\!\!\!\!- \frac{{\rm d}p'}
{2\pi}
\Theta(|p'|-p_f)\frac{
\sin \left(\theta(p)-\theta(p')\right)}{(p-p')^2} = 0 \ ,
\quad (|p|> p_f) \ ,
\label{d27}
\end{equation}
whereas the condensate now becomes
\begin{equation}
\langle \bar{\psi}\psi \rangle = -N \int \frac{{\rm d}p}{2\pi} \theta(|p|-p_f)
\sin \theta(p) \ .
\label{d28}
\end{equation}
The gap equation (\ref{d27}) can easily be solved numerically for various $p_f$.
The resulting condensate is shown in Fig.~3.
We find that it decreases monotonically with increasing
density, disappearing at a critical Fermi momentum
\begin{equation}
p_f^c \approx 0.117 \left( \frac{Ng^2}{2\pi} \right)^{1/2} \ .
\label{d28a}
\end{equation}
This behaviour is strikingly similar to the
corresponding result for the Gross-Neveu model depicted in Fig.~1, again
suggesting
some phase transition with restoration of chiral symmetry at high
density. Here we are not able to go on and discuss
whether we are
dealing with a first or second order phase transition. The reason
lies in the following problem:
If we compute the energy density (\ref{d26}) for the 't~Hooft model we
discover that subtraction of the value at $p_f=0$ is not sufficient to
give a finite result. Unlike in the Gross-Neveu model,
the difference is still IR divergent.
To be able to proceed, we enclose the system once more in a box of length
$L$. We then find that the divergence is due to the last term
in Eq.~(\ref{d26}) (the one which does not involve the Bogoliubov angles)
which now contributes the following double sum to the energy per color,
\begin{equation}
\left. \frac{E}{N}\right|_{\rm div} = \frac{Ng^2 L}{16 \pi^2} \sum_{p\in I}
\sum_{n\neq 0,(p-n)\not\in I} \frac{1}{n^2} \ .
\label{d29}
\end{equation}
Here antiperiodic boundary conditions for fermions have been
employed in the box regularization, and correspondingly
the interval $I$ is defined in the following way,
\begin{equation}
I = [-n_f, n_f] \ \ \ \mbox{for} \ \ B=2 n_f+1 \ \mbox{odd} \ , \ \ \
I = [-n_f-1, n_f] \ \ \ \mbox{for} \ \ B=2n_f+2\  \mbox{even} \ .
\label{d29a}
\end{equation}
The result (\ref{d29}) is even more alarming than the non-convex behaviour of
${\cal E}$ in the Gross-Neveu model, Fig.~2, due to its $L$-dependence.
Adding quarks to the
vacuum causes the energy to increase by an infinite amount in the limit
$L \to \infty$. Evaluating the double sums in Eq.~(\ref{d29}) for
low values of $B$, we obtain information on the origin of this divergent
behaviour. For $B=1$
($I=[0,0]$) in particular, the calculated baryon mass (to leading order in $L$) is
\begin{equation}
M_B= N \left(\frac{N g^2 L}{48}\right) \ .
\label{d30}
\end{equation}
This is the same relation as $M_B=N m_0$ in the Gross-Neveu model
except that the physical fermion mass is replaced by the infinite
constant $Ng^2 L/48$ characteristic of confinement, cf. Eq.~(\ref{d24}).
For larger values of
$B$ Eq.~(\ref{d29}) does not simply yield multiples of the baryon mass
(\ref{d30}), but one finds
``interaction effects" of the same order of magnitude as the mass.
As far as $N$-counting is concerned, this is still in agreement with
Witten's analysis of baryons at large $N$ \cite{Witten79}. However,
since these delocalized baryons are presumably not very physical in the
't~Hooft model, we refrain from further discussing these effects.

Summarizing,
the problems encountered in the Gross-Neveu model with translationally
invariant baryonic matter again show up in the 't~Hooft model, although in
a much more severe form.
The physics reason is clear: In the Gross-Neveu
model the cost of distributing $N$ fermions over the whole space is
governed by their physical mass; in the 't~Hooft model, due to confinement
of quarks, the corresponding quark effective mass diverges with the
volume. On the other
hand, it is known that both models
do possess massless, delocalized baryons in the chiral limit. Evidently, this
has to be accounted for, and we conclude that the naive, translationally
invariant Hartree-Fock approximation fails miserably in describing
the properties of baryonic matter.

\section*{4) Massless baryons and baryonic matter in the Skyrme picture}

The existence of massless baryons in the chiral limit of the
't~Hooft model has been demonstrated by
bosonization \cite{Affleck,Abdalla}, variational \cite{Salcedo}, and
light-cone \cite{Lenz}
techniques. These exotic objects are characteristic for 1+1 dimensional
models with broken chiral symmetry and, as such, also present in the
chiral Gross-Neveu model. A particularly illuminating
derivation is due to Salcedo {\em et al.} \cite{Salcedo}.
These authors point out
that the potential energy in such models is invariant under local
chiral transformations, unlike the kinetic term which is
only invariant under global ones. This led them to the following
variational ansatz for the one-body density matrix of the baryon,
\begin{equation}
\rho(x,y)={\rm e}^{{\rm i}\chi(x)\gamma_5}\rho_{\rm v}(x-y)
{\rm e}^{-{\rm i}\chi(y)\gamma_5} \ .
\label{d30a}
\end{equation}
Here $\rho_{\rm v}(x-y)$ is the vacuum density matrix.
If the vacuum breaks chiral symmetry, one can generate with
expression (\ref{d30a})
a new (exact or approximate) Hartree-Fock solution which breaks
translational invariance but can carry non-zero baryon number.
As shown in \cite{Salcedo}, the baryon density is given by
\begin{equation}
\rho_B(x)={\rm tr}(\rho(x,x)-\rho_{\rm v}(0))=
  \frac{1}{\pi}\partial_x \chi(x) \ ,
\label{d30b}
\end{equation}
so that
the baryon number coincides with the winding number of the chiral phase
$\chi(x)$,
\begin{equation}
B=\int_0^L {\rm d}x \frac{1}{\pi}\partial_x \chi = \frac{\chi(L)-\chi(0)}{\pi}
\in  \ZZ  \ .
\label{d31}
\end{equation}
(Notice that $\chi(L)-
\chi(0)$ must be an integer multiple of $\pi$ since otherwise bilinear
fermion observables would no longer be periodic.)
For this topological reasoning it is again recommendable to work in a
finite box of size $L$. The topological interpretation of the baryon
number also agrees with exact results of Ref.~\cite{Lenz}  which
were not restricted to the large $N$ limit.
In the absence of an explicit quark mass term,
the ground state energy obtained from Eq.~(\ref{d30a}) is
\begin{equation}
E[\rho] = E[\rho_{\rm v}] + N \int_0^L {\rm d}x \frac{1}{2\pi}
(\partial_x\chi)^2 \ .
\label{d32}
\end{equation}
This result holds independently of the specific model,
since the potential energy does not contribute to $E[\rho]-E[\rho_{\rm v}]$.
Differences between various
models are of course
still present in the vacuum density matrix $\rho_{\rm v}$
in Eq.~(\ref{d30a}) but do not manifest themselves in the baryon energy.
Minimizing $E[\rho]$ with respect to $\chi$ yields the free (static)
bosonic equation
\begin{equation}
\partial_x^2\chi(x)=0 \ , \qquad \chi(L)= \chi(0)+ \pi B
\label{d33}
\end{equation}
with the solution
\begin{equation}
\chi(x)=\pi B \left(\frac{x-x_0}{L}\right) \ .
\label{d34}
\end{equation}
Here
$x_0$ is a parameter which reflects the breakdown of translational
invariance.
The baryon density is $x$-independent ($\rho_B(x)=B/L$) as follows
more generally from axial current conservation in the chiral limit
\cite{Salcedo}.
However, the scalar and pseudoscalar condensates acquire a non-trivial
$x$-dependence,
\begin{eqnarray}
\langle \bar{\psi}\psi\rangle  & = & \langle \bar{\psi}\psi \rangle_{\rm v}
\cos \left( 2 \pi B (x-x_0)/L\right) \ , \nonumber \\
\langle \bar{\psi}{\rm i}\gamma_5 \psi\rangle  & = &
-\langle \bar{\psi}\psi \rangle_{\rm v}
\sin \left( 2 \pi B (x-x_0)/L\right) \ .
\label{d35}
\end{eqnarray}
Since fluctuations of $\chi(x)$ describe the massless Goldstone boson field,
the baryon picture emerging here is very similar in spirit to the Skyrme model
\cite{Skyrme}. The fact that the baryon is a topological soliton will become
somewhat clearer once we include a small bare quark mass (see Sect.~5)
but this solitonic character also holds in the strict chiral limit
considered here.

We can now discuss the baryon as well as baryonic matter from this point of
view. The single baryon ($B=1$) is spread out over the whole space,
the chiral phase $\chi(x)$ making one turn with constant speed to minimize
the kinetic energy (Fig.~4). The baryon energy is,
using Eqs.~(\ref{d32}-\ref{d34}),
\begin{equation}
E_B=  N \frac{\pi}{2L}  \ .
\label{d36}
\end{equation}
This confirms that indeed the baryon becomes massless in the limit
$L\to \infty$. Incidentally, expression (\ref{d36}) is identical to the kinetic
energy of $N$ non-interacting, massless quarks in the lowest momentum
state available for antiperiodic boundary conditions. Nevertheless,
we are not dealing with the free, chirally symmetric theory, but with
the broken phase of an interacting theory where the quarks are
massive or even confined.

It may be worthwhile to contemplate the structure of the baryon for a
moment from the point of view of the relativistic Hartree-Fock
approximation.
In chirally non-invariant models one would suspect that the baryon
comprises the filled Dirac sea plus one filled, positive energy valence
level. This
is exactly what one finds analytically in the non-chiral Gross-Neveu
model \cite{Pausch}, or numerically
in QCD$_2$ with heavy quarks \cite{Salcedo}. The picture implied by
the ansatz (\ref{d30a}) in the chiral limit is rather different though.
Denoting the negative energy
single particle orbitals in the Dirac sea by $\varphi_k^{(-)}(x)$
(solutions of the first quantized Dirac equation with Hartree-Fock potential),
the Skyrme type baryon (\ref{d30a}) admits the density matrix
\begin{equation}
\rho(x,y)={\rm e}^{{\rm i} \pi x \gamma_5/L} \sum_k \varphi_k^{(-)}(x)
\varphi_k^{(-)\dagger}(y) {\rm e}^{-{\rm i} \pi y \gamma_5/L}
\label{d37}
\end{equation}
where
\begin{equation}
\varphi_k^{(-)}(x)={\rm e}^{{\rm i}kx} v(k) \ ,
\label{d38}
\end{equation}
and the $k$ are discrete momenta appropriate to the interval of length $L$.
We first observe that the chiral phase factor splits the
momenta of the right- and left-handed components into $k\pm \pi/L$.
Since the transformed single particle wave functions are no longer
momentum eigenstates, translational invariance is lost.
Secondly, we note that the presence of an infinite Dirac sea is
crucial for getting the extra baryon charge, rather than a single
valence state.
If the sum over occupied states $k$ in Eq.~(\ref{d37}) was finite,
we would trivially conclude that $\rho(x,y)$ and $\rho_{\rm v}(x-y)$
belong to
the same baryon density ($\rho_B(x)={\rm tr}\rho(x,x)$). Due to the
infinite number of occupied states however,
$\rho_{\rm v}(x-y)$ develops a singularity at $x=y$, and one has to do
a more careful point splitting in order to compute the baryon density.
The divergence is due to the UV region and therefore determined by the
free theory
(for more details, cf. Ref.~\cite{Salcedo}),
\begin{equation}
\lim_{x\to y} {\rm tr}(\rho(x,y)-\rho_{\rm v}(x-y)) = \lim_{z\to 0}
{\rm tr} \left\{ {\rm e}^{{\rm i}\pi z \gamma_5/L}
\left(\frac{1}{2}\delta(z)
-\frac{{\rm i}\gamma_5}{2\pi z}\right)-\frac{1}{2}\delta(z) \right\}
  =    \frac{1}{L} \ .
\label{d40}
\end{equation}
The result $1/L$ is the baryon density for $B=1$.
This mechanism is strongly reminiscent of the calculation of anomalous current
commutators, for instance in the Schwinger model. The extra baryon number
does not reside in a valence level added on top of the Dirac sea
but somehow emerges from the
bottom of the Dirac sea if one modifies all the levels slightly ---
it is a vacuum polarization effect.

Equipped with this exotic kind of baryon, we can now easily find the
ground state of the system for any baryon density.
As discussed above and illustrated in Fig.~4,
the single baryon consists of one turn of a ``chiral spiral" (parametrized
by $\chi(x)$) over the total spatial length $L$ of the system --- admittedly
a somewhat elusive object in the thermodynamic limit.
A finite density $\rho_B=B/L=p_f/\pi$ on the other hand implies that
\begin{equation}
\chi(x)=p_f (x-x_0) \ ,
\label{d42}
\end{equation}
{\em i.e.}, one full rotation over a physical distance which has a well
defined
limit for $L\to \infty$, namely 2/$\rho_B$. The baryon density remains
constant in space,
but the condensates are modulated as
\begin{eqnarray}
\langle \bar{\psi} \psi\rangle  & = & \langle \bar{\psi} \psi \rangle_{\rm v}
\cos 2 p_f (x-x_0) \ , \nonumber \\
\langle \bar{\psi}{\rm i}\gamma_5 \psi\rangle  & = &
-\langle \bar{\psi} \psi \rangle_{\rm v}
\sin 2 p_f (x-x_0) \ .
\label{d43}
\end{eqnarray}
They can be viewed as projections of a ``chiral spiral" of radius
$|\langle \bar{\psi}\psi \rangle_{\rm v} |$ onto two
orthogonal planes, see Fig.~5.
This state breaks translational symmetry; it is a crystal. In fact, it
may be viewed as the simplest possible realization of the old idea
of a Skyrme crystal \cite{Klebanov}, here in the context of large $N$
two-dimensional field
theories. One cannot tell where one baryon begins and ends ---  each full turn
of the spiral contains baryon number 1. Only the condensates reveal that
translational symmetry has been broken down to a discrete subgroup. The
energy density
of this unusual kind of ``nuclear matter" is simply (after subtracting
the vacuum energy density)
\begin{equation}
\frac{{\cal E}}{N}=\frac{p_f^2}{2\pi} \ .
\label{d44}
\end{equation}
Surprisingly,
this is exactly what one would expect for a free Fermi gas of
massless quarks although Eq.~(\ref{d44}) holds for interacting
theories where the vacuum has lower energy due to chiral symmetry
breaking. In Fig.~6 we compare the energy density for this state to the
ones discussed
above for the Gross-Neveu model, where translational symmetry had been
assumed. The crystal is always energetically favored,
the dependence on $p_f$ is now
convex, and there is no trace of a phase transition, neither first
nor second order, at any density. The horizontal slope at $p_f=0$
correctly
signals the presence of massless baryons and eliminates the above
mentioned problems with the spurious massive, delocalized baryons.
We cannot even draw the corresponding picture for the 't~Hooft model,
simply because in this case the quark Fermi gas is infinitely
higher in energy than the Skyrme crystal for $L\to \infty$.
Nevertheless, all the results for baryonic matter discussed
in this section apply to the 't~Hooft model as well.

In the high density limit the oscillations
of the condensates become more and more rapid. If we are interested only
in length scales large as compared to $1/p_f$, the
condensates average to zero. In this sense, one might argue that
chiral symmetry gets restored at high density,
although not in the naive way suggested by Fig.~2.

Finally, we remark that the ``chiral spiral" ground
state for fixed baryon density still preserves one continuous, unbroken
symmetry,
namely the combination of translation and chiral rotation generated by
$P+p_f Q_5$ ($P$: momentum operator, $Q_5$: axial charge).
One would therefore predict that RPA excitations on this ground
state \cite{Salcedo,Lenz} (or mesons in nuclear matter) will have only one
collective, gapless mode,
a hybrid of a ``phonon" and a ``pion".

\section*{5) Non-vanishing bare quark masses}

In Ref.~\cite{Salcedo} the Skyrme picture of the baryons
in the 't~Hooft model and chiral Gross-Neveu model was developed for
small, finite bare quark masses, using the expression in Eq.~(\ref{d30a})
as a variational ansatz. For a single baryon, these authors have tested the
accuracy of their procedure against the full, numerical Hartree-Fock
calculation on a lattice. The results agreed perfectly
at $m_q=0.05$ and were still surprisingly good at $m_q=0.20$, in units of
$\sqrt{Ng^2/2\pi}$. This makes it very tempting to speculate that the
corresponding variational calculation can also give us a reliable
picture of baryonic matter at finite density, away from the chiral limit.
As compared to the formulae in the preceding section, the only change is
the fact that the bare mass term now also contributes to the energy
functional Eq.~(\ref{d32}),
\begin{equation}
E[\rho] = E[\rho_{\rm v}]+ N \int_0^L {\rm d}x \left\{ \frac{1}{2\pi}
\left(\partial_x \chi\right)^2+ \frac{m_q \langle \bar{\psi}\psi
\rangle_{\rm v}}{N} \left( \cos 2\chi-1 \right) \right\} \ .
\label{d45}
\end{equation}
Here the condensate $\langle \bar{\psi}\psi \rangle_{\rm v}$ refers to
the vacuum in the chiral limit.
Varying with respect to
$\chi(x)$ then gives the static sine-Gordon equation \cite{Scott},
\begin{equation}
\partial_x^2 \chi + \frac{2\pi m_q \langle \bar{\psi}\psi \rangle_{\rm v}}
{N} \sin 2 \chi  = 0 \ ,
\label{d46}
\end{equation}
from which one reads off the ``pion" mass (two dimensional version
of the Gell Mann-Oakes-Renner relation \cite{Oakes})
\begin{equation}
m_{\pi}^2 = - 4 \pi m_q \frac{\langle \bar{\psi}\psi\rangle_{\rm v}}{N} \ .
\label{d47}
\end{equation}
The $B=1$ baryon can be identified with the familiar kink solution
of the sine-Gordon equation,
\begin{equation}
\chi(x)= 2 \arctan \left({\rm e}^{m_{\pi}(x-x_0)}\right) \ ,
\label{d48}
\end{equation}
with mass
\begin{equation}
M_B=N \frac{2 m_{\pi}}{\pi} \ .
\label{d49}
\end{equation}
Since the single baryon has been discussed in detail in Ref.~\cite{Salcedo}, let us
immediately
turn to multi-kink solutions as candidates for baryonic matter.
Luckily, the sine-Gordon kink crystal has already been studied
thoroughly
in the literature, first in solid state physics \cite{McMillan,Theodorou}
and more recently as a toy model for the Skyrme crystal \cite{Takayama},
in terms of Jacobi elliptic functions and elliptic integrals
\cite{Abramovitz}.
We take over the results from Ref.~\cite{Takayama}
which is close in spirit to the present study although the authors did not
have in mind two-dimensional large $N$ field theories.
Adapting the formulae of this work to our notation, the following
steps allow us to generalize the Skyrme crystal of the previous
section to small, finite bare quark masses: Let $m_{\pi}$ denote the
mass of the Goldstone boson, Eq.~(\ref{d47}), and $\bar{\rho}_B=
p_f/\pi$ the average baryon density (this is our definition of $p_f$
for the case of broken translational symmetry).
We then first have to solve the
transcendental equation
\begin{equation}
\frac{\pi m_{\pi}}{p_f} = 2 k {\bf K}(k)
\label{d50}
\end{equation}
for $k$ where ${\bf K}(k)$ is the complete elliptic integral of the
first kind. The sine-Gordon kink crystal is then given by the following
solution of Eq.~(\ref{d46}),
\begin{equation}
\chi(x) = \frac{\pi}{2} + {\rm am}(\xi,k) \ , \qquad \xi=
\frac{m_{\pi}}{k}(x-x_0) \ ,
\label{d51}
\end{equation}
(${\rm am}(\xi,k)$ is the Jacobian elliptic amplitude function). From this,
we can
express the baryon density and the various condensates in terms of
further Jacobian elliptic
functions (${\rm dn}, {\rm sn}, {\rm cn}$) as
follows,
\begin{eqnarray}
\rho_B(x)&=& \frac{1}{\pi} \partial_x \chi(x) \ = \
\frac{m_{\pi}}{\pi k} {\rm dn}(\xi,k) \ ,
\nonumber \\
\langle \bar{\psi}\psi \rangle & = & \langle \bar{\psi}\psi \rangle_{\rm v}
\cos 2 \chi(x) \ = \
- \langle \bar{\psi}\psi \rangle_{\rm v}
\left( {\rm cn}^2(\xi,k)-{\rm sn}^2(\xi,k)\right) \ ,
\nonumber \\
\langle \bar{\psi}{\rm i}\gamma_5 \psi \rangle & = &
-\langle \bar{\psi}\psi \rangle_{\rm v} \sin 2 \chi(x) \ = \
 \langle \bar{\psi}\psi \rangle_{\rm v}
2 {\rm sn}(\xi,k) {\rm cn}(\xi,k)\ .
\label{d52}
\end{eqnarray}
Here $\xi$ is as defined in Eq.~(\ref{d51}). Finally, the energy divided
by the volume of this kind of matter is given by
\begin{equation}
\frac{{\cal E}}{N}= \frac{m_{\pi} p_f}{4 \pi^2} \left\{ \frac{8}{k} {\bf E}
(k) + 4 k \left(1-\frac{1}{k^2}\right) {\bf K}(k) \right\} \ ,
\label{d53}
\end{equation}
${\bf E}(k)$ denoting the complete elliptic integral of the second kind.

Let us now illustrate these results in two regimes of interest, namely
at low and high density. At low density ($p_f\ll m_{\pi}$) $k$ in
Eq.~(\ref{d50}) approaches 1 exponentially, and the baryon density
features a chain of well resolved lumps whose shape is determined by
the single kink solution (Fig.~7). Likewise, the condensates behave
like those of a
single baryon: $\langle \bar{\psi}\psi \rangle$ changes
from the vacuum value outside the baryons to its negative in their center
whereas $\langle \bar{\psi}{\rm i}\gamma_5 \psi \rangle $ is peaked in the
surface region of each baryon (Figs. 8, 9). These condensates
are projections of the distorted ``chiral spiral" shown in Fig.~10.
The energy (\ref{d53}) for low densities behaves as
\begin{equation}
{\cal E} \approx N \frac{2 m_{\pi}p_f}{\pi^2} = M_B \rho_B \ ,
\label{d54}
\end{equation}
showing the expected connection to the baryon mass. At high
densities ($p_f \gg m_{\pi}$), $k$ approaches 0 like
\begin{equation}
k \approx \frac{m_{\pi}}{p_f} \ .
\label{d55}
\end{equation}
Thus $\xi$ in Eq.~(\ref{d51}) becomes $p_f (x-x_0)$. Moreover, for $k\to 0$,
the
Jacobian elliptic functions ${\rm am}(\xi,k), {\rm sn}(\xi,k),{\rm  cn}(\xi,k)$
are known to reduce to the argument $\xi$ and the ordinary trigonometric
functions $\sin \xi$ and $\cos \xi$, respectively. We thus recover the
results for the simple chiral spiral in Sect.~4
(the parameter $x_0$ has to be readjusted to take care of the shift by $\pi/2$
in Eq.~(\ref{d51})). The energy in this case is approximately
\begin{equation}
\frac{{\cal E}}{N} \approx \frac{p_f^2}{2\pi} + \frac{m_{\pi}^2}{8\pi} \ .
\label{d56}
\end{equation}
The condensates look very much like the sin- and cos-functions of
the massless case and need
not be plotted. The baryon density wiggles around a constant value,
reflecting the strong overlap of the baryons, and can be approximated
at high density by
\begin{equation}
\rho_B(x) \approx \frac{p_f}{\pi} \left( 1-\frac{1}{2} \left( \frac{m_{\pi}}
{p_f}\right)^2 \sin^2p_f(x-x_0)\right) \ .
\label{d57}
\end{equation}
The behaviour of the baryon density $\rho_B(x)$ as one increases
$p_f$ ({\em i.e.}, the mean density) is
illustrated in Fig.~11. In the chiral- or high-density
limit ($m_{\pi}/p_f \to 0$) $\rho_B(x)$ eventually becomes $x$-independent.
This provides us with another way of understanding the structure of
matter described
in the previous section, namely as arising from a chain
of very extended, strongly overlapping lumps.

Finally, let us come back to the question of validity of the variational
calculation based on the chirally modulated vacuum density matrix
(\ref{d30a}), which we have left open so far.
At very low densities when the individual baryons are
far apart, we can presumably rely on the numerical
results of Ref.~\cite{Salcedo}
since the interaction effects between the baryons
become small (as discussed in Refs.~\cite{Salcedo} and \cite{Takayama},
the baryon-baryon interaction is
repulsive and falls off exponentially with the pion Compton wavelength).
At high densities, on the other hand, one would expect
that a finite quark mass cannot make much difference as long as
$p_f \gg m_{\pi}$.
This takes us back to
the massless case discussed in Sect.~4. In this
limit, in turn,
it is easy to convince oneself that the calculation
becomes exact in the sense that one gets a true solution of the
Hartree-Fock equation. Thus for instance for the 't~Hooft model,
the massless Hartree-Fock equation reads \cite{Salcedo}
\begin{equation}
\omega_n \varphi_{\alpha}^{(n)}(x) = -{\rm i}(\gamma_5)_{\alpha \beta}
\frac{\partial}{\partial x}
\varphi_{\beta}^{(n)}(x) + \frac{Ng^2}{4} \int {\rm d}y |x-y|
\rho_{\alpha \beta}(x,y) \varphi_{\beta}^{(n)}(y) \ .
\label{d58}
\end{equation}
Upon substituting
\begin{equation}
\varphi_{\alpha}^{(n)}(x) = \left({\rm e}^{-{\rm i} p_f x \gamma_5}
\right)_{\alpha \beta}
\tilde{\varphi}^{(n)}_{\beta}(x)
\label{d59}
\end{equation}
as we are instructed to do by the ansatz (\ref{d30a}), we discover
that $\tilde{\varphi}^{(n)}$ does indeed solve the Hartree-Fock equation,
the only change being that the single particle energy $\omega_n$
gets replaced by $\omega_n + p_f$. The same argument goes through
in the chiral Gross-Neveu model, or in any field theory where the
interaction term has a local chiral invariance.
This proves that the result becomes exact in
the chiral limit (to leading order in the $1/N$ expansion, of course)
and makes plausible the hypothesis that it also correctly describes
the high density regime for
finite quark masses as long as $p_f\gg m_{\pi}$.

\section*{6) Summary and conclusions}

In this paper, we have addressed the problem of baryonic matter
in a certain class of exactly soluble field theoretic models, namely
chirally invariant, large $N$, interacting fermion theories.
We started out from a seemingly innocuous and well understood problem,
the chiral Gross-Neveu model at finite density and identified
one remaining weak spot: The energy density for
baryonic matter in the standard Hartree-Fock approach does
not have the correct low density limit which can be predicted from the
known baryon spectrum of the theory.
The origin of this problem which
is not cured by a mixed phase approach conceptionally related
to the bag model, is evidently the assumption of translational
invariance. Whereas this inconsistency can perhaps be ignored in the
Gross-Neveu
model (as it has been so far, to the best of our knowledge), in QCD$_2$,
it becomes fatal: Due to confinement, the
analogous calculation yields an infinite energy for delocalized
baryons or quark matter. This is unavoidable
if one is careful in treating the infrared behaviour of
the quark single particle energies. These findings have prompted us to
think more thoroughly about the structure of baryons in such models and
possible implications for the matter problem. We found that it
takes only very little effort to generalize a previous Skyrme type
treatment of the single, massless baryon to the case of baryonic matter.
In the chiral
limit, an extremely simple, yet non-trivial, picture emerges:
Both the baryon and dense matter are described by a spatially varying
chiral angle which is best characterized as a ``chiral spiral" with
constant
helix angle. The number of windings within
the full space of length $L$ measures the number of baryons in the box.
Since, by construction,  this kind of state
does not cost any potential energy in addition to what is already
stored in the vacuum, one does not have to pay the expected high price
for delocalizing quarks. The energy density of baryonic matter
is identical to that of a free Fermi gas of massless quarks,
in spite of the presence of
interaction effects. The same picture applies
to the chiral Gross-Neveu as well as to the massless 't~Hooft model and
should be generic for all chiral large $N$ models. The
baryon density is constant in space as a consequence of axial
current conservation, and we have verified that the whole scenario is exact
to leading order in the $1/N$ expansion.
In a slightly more speculative vein we then investigated
modifications due to a small bare quark mass. Here our task
was greatly facilitated by the fact that we only needed to pull
together two independent investigations, the one of Ref.~\cite{Salcedo}
of the single baryon in field theoretic models with the one of
Ref.~\cite{Takayama} of the sine-Gordon kink crystal, both inspired
in some way by Skyrme's original ideas. As a result, we have
arrived at a rather comprehensive picture of matter at low and
high density on the scale of the pion Compton wavelength.
The crystal structure now becomes more conspicuous since also the baryon
density displays a lattice of individual lumps.
As an additional
bonus, we have obtained a purely classical, mechanical model of what is
going on (the sine-Gordon equation describes a chain of coupled pendulums, the
quark mass playing the role of gravity).
Given our starting point, namely
the problem of baryonic matter in two dimensional quantum field theories
like the Nambu--Jona-Lasinio model or QCD, this is rather amusing.

It is noteworthy that a similar chiral structure of fermionic matter
has been reported previously in a variety of models different from
the present ones. This indicates that the basic results are more generally
valid than our derivation might suggest. We mention here in
particular
the early work on the massive Schwinger model \cite{Fischler79}
and the more recent work on
the massless Schwinger model with inert background charge
\cite{Kao94} and QCD$_2$ with a
finite number of colors and flavours \cite{Christiansen96}. Even more
surprising are perhaps quite a number of speculations about
spatially inhomogeneous chiral
condensates with the same wave number as in our case, but
in 3+1 dimensions, in the context of pion
condensation \cite{Kutschera90}, large $N$ QCD
\cite{Deryagin92,Shuster99,Park99}, or effective chiral models
\cite{Sadzikowski00}. In some of these works, the analogy with the
Overhauser effect
and spin-density waves (pairing of particle holes on opposite sides
of the Fermi sphere) has been stressed. Although the language used is
quite different from ours, there is no doubt that we are dealing
with the same physical phenomenon.

As a last remark, we wish to comment on the original Gross-Neveu model with
only discrete chiral symmetry (pure $(\bar{\psi}\psi)^2$-interaction). Most
of the studies of the phase diagram for the Gross-Neveu model have in fact been
performed for this model,
and one might think that our analysis does not have anything to say about
it. However, the criticism of Sect.~2 also applies here.
Since the non-chiral Gross-Neveu model has (massive) bound baryons, the low
density behaviour of the energy
obtained in standard Hartree-Fock approximation
cannot be correct, and the phase diagram may also have to be reconsidered.

\vskip 1.0cm
We thank A.C. Kalloniatis for a critical reading of the manuscript.
Partial support of this work by the Bundesministerium f\"ur Bildung,
Wissenschaft, Forschung und Technologie is gratefully acknowledged.

\section*{Appendix: Hartree-Fock single particle energies
in the 't~Hooft model}

The Hamiltonian for the massless 't~Hooft model in the axial gauge
has the form
\begin{equation}
H = \sum_{p,i} \frac{2\pi}{L}(p+1/2)\left( a_i^{\dagger}(p)a_i(p)
-b_i^{\dagger}(p)b_i(p) \right) + \frac{g^2 L}{16 \pi^2}
\sum_{ij,n\neq 0} \frac{j_{ij}(n)j_{ji}(-n)}{n^2} \ .
\label{A1}
\end{equation}
Here we have regularized the theory by enclosing it in a box of
length $L$ with antiperiodic boundary conditions for the fermions.
Note that this form is only valid in the limit $L\to \infty$
\cite{Schoen}. The $a_i(p), b_i(p)$ denote right- and left-handed quark
operators, respectively, and the currents $j_{ij}(n)$ can be taken
in the U($N$) form at large $N$,
\begin{equation}
j_{ij}(n) = \sum_p \left( a_j^{\dagger}(p)a_i(p+n)+b_j^{\dagger}(p)
b_i(p+n) \right) \ .
\label{A2}
\end{equation}
It is important to understand that the Coulomb term still
contains one- and two-body operators which can be disentangled by
normal-ordering (up to $1/N$ corrections) as follows,
\begin{eqnarray}
\sum_{ij}j_{ij}(n)j_{ji}(-n) & = & N \sum_{i,p} \left(a_i^{\dagger}(p)
a_i(p) + b_i^{\dagger}(p)b_i(p) \right)  \\
& - & \sum_{ij,pq} \left(
a_j^{\dagger}(p)a_j(q) a_i^{\dagger}(q+n)a_i(p+n) +
a_j^{\dagger}(p)b_j(q) b_i^{\dagger}(q+n)a_i(p+n)   \right.
\nonumber \\
& &  \left.
+ b_j^{\dagger}(p)a_j(q) a_i^{\dagger}(q+n)b_i(p+n) +
b_j^{\dagger}(p)b_j(q) b_i^{\dagger}(q+n)b_i(p+n) \right) \nonumber
\label{A3}
\end{eqnarray}
The Hamiltonian can be decomposed correspondingly into one- and
two-body operators. Using the basic vacuum expectation values
\begin{eqnarray}
\sum_i \langle 0 | a_i^{\dagger}(p) a_i(q)|0 \rangle &=&
\frac{N}{2}\delta_{pq}(1-
\cos \theta(p) )\ , \nonumber \\
\sum_i \langle 0 | b_i^{\dagger}(p) b_i(q) | 0 \rangle &=&
\frac{N}{2}\delta_{pq}(1+
\cos \theta(p) )\ , \nonumber \\
\sum_i \langle 0 | a_i^{\dagger}(p) b_i(q) | 0 \rangle &=&
\sum_i \langle 0 | b_i^{\dagger}(p) a_i(q) | 0 \rangle \ = \
-\frac{N}{2}\delta_{pq}  \sin \theta(p) \ ,
\label{A4}
\end{eqnarray}
the vacuum expectation value of these two contributions are found to be
\begin{eqnarray}
\langle 0 |H^{(1)}|0\rangle & = & N\sum_p \left(
-\frac{2\pi}{L}\left(p+\frac{1}{2}\right)
\cos \theta(p) +\frac{N g^2 L}{48} \right) \ ,
\nonumber \\
\langle 0 |H^{(2)}|0\rangle & = & N\sum_p \left( -\frac{N g^2 L}
{32 \pi^2} \sum_{n \neq 0} \frac{1}{n^2} \left[ 1 + \cos (\theta(p)
-\theta(p+n)) \right]\right) \ .
\label{A5}
\end{eqnarray}
The $N g^2 L/48$ term in the 1-body part is just minus twice the ``1"
term in the 2-body part (remember that $\sum_{n\neq 0} 1/n^2
= \pi^2/3$). Hence, in the sum of both terms, the Coulomb energy
involves the combination
\begin{equation}
\frac{1}{n^2} [\cos(\theta(p)-\theta(p+n))-1]
\label{A6}
\end{equation}
where the infrared divergence has been tamed since denominator and
numerator both vanish at $n=0$. This cancellation
between quark self-energy and Coulomb potential is similar to what
happens in the meson equation of the 't~Hooft model. The continuum limit
then yields Eq.~(\ref{d20}). It is important to distinguish between
one- and two-body operators
here, because they enter with different relative weights in the single particle
energies and in the total energy. Indeed, in the Hartree-Fock approach,
if the
single particle energies are decomposed according to their 1- and 2-body
contributions as
\begin{equation}
\omega(p) = \omega^{(1)}(p) + \omega^{(2)}(p) \ ,
\label{A7}
\end{equation}
then the ground state energy is
\begin{equation}
\langle 0 | H | 0 \rangle = N \sum_p \left( \omega^{(1)}(p)
+ \frac{1}{2} \omega^{(2)}(p)  \right)\ .
\label{A8}
\end{equation}
The factor 1/2 is necessary to avoid double counting of the 2-body
interaction term.
By comparison with Eq.~(\ref{A5}), we can turn this observation around
and simply read off
the single particle energies. We find in this way
\begin{eqnarray}
\omega^{(1)}(p) & =& -\frac{2\pi}{L}\left(p+\frac{1}{2}\right) \cos
\theta(p) +\frac{N g^2 L}{48}  \ ,
\nonumber \\
\omega^{(2)}(p) & = &  -\frac{N g^2 L}
{16 \pi^2} \sum_{n \neq 0} \frac{1}{n^2} \left( 1 + \cos (\theta(p)
-\theta(p+n)) \right) \ .
\label{A9}
\end{eqnarray}
Adding up the two contributions to $\omega(p)$, the ``1" term
is now cancelled instead of changing sign.
This is the reason why in the continuum limit we get the
badly infrared divergent expression (\ref{d23}) for the quark energies.
This short-cut derivation gives the same result as the more elaborate
approach of Ref.~\cite{Lenz}, where a single particle Hartree-Fock
Hamiltonian was first identified by commuting $H$ with the quark operators
and subsequently diagonalized.

\newpage
\subsection*{Figure captions}
\begin{enumerate}
\item
Physical fermion mass as a function of the Fermi momentum in
the Gross-Neveu model, in units of $m_0$.
\item
Energy density per color as a function of the Fermi momentum in the
Gross-Neveu model. Solid curve: Chirally symmetric solution ($m=0$);
diamonds: Broken chiral symmetry ($m$ according to Fig.~1);
dashed straight line: Mixed phase. Units of $m_0$.
\item
Quark condensate as a function of the Fermi momentum in the
't~Hooft model, in units of $(Ng^2/2\pi)^{1/2}$.
\item
The complex condensate $\langle \bar{\psi}\psi\rangle
+{\rm i} \langle \bar{\psi} {\rm i}\gamma_5 \psi \rangle$
for the single baryon in units of
$\langle \bar{\psi}\psi \rangle_{\rm v}$, as a function
of $x$ in units of $L$ (chiral limit).
\item
Same as Fig.~4, but for baryonic matter. Each full turn of the
spiral increases the baryon number by one unit.
\item
Same as Fig.~2 (Gross-Neveu model). Here we have included the energy
density of the Skyrme crystal type of state (crosses), the true
ground state.
\item
Solid curve:
Spatial oscillation of the baryon density in the regime $p_f \ll m_{\pi}$;
circles: Baryon density for a single baryon.
\item
Solid curve: Spatial oscillations of the scalar chiral condensate
in the regime $p_f \ll m_{\pi}$; circles: Scalar chiral condensate
for a single baryon.
\item
Same as Fig.~8, but for the pseudoscalar chiral condensate.
\item
Illustration of the distorted ``chiral spiral" for baryonic
matter at non-zero bare quark mass.
\item
Spatial dependence of baryon density as it evolves with
increasing average density (or Fermi momentum), in units of
$m_{\pi}$.
\end{enumerate}

\newpage

\setlength\parindent{0cm}
\setlength\parskip{0pt}

\setlength\textheight{25cm}

\begin{figure}
\begin{center}
\begin{psfrags}
\psfrag{Fig. 1}[c][c]{}
\psfrag{pf}[t][b]{$p_f$}
\psfrag{piabc}[c][l]{$\displaystyle {m}$}
\psfrag{0}[c][c]{$0$}
\psfrag{0.1}[c][c]{$0.1$}
\psfrag{0.2}[c][c]{$0.2$}
\psfrag{0.3}[c][c]{$0.3$}
\psfrag{0.4}[c][c]{$0.4$}
\psfrag{0.5}[c][c]{$0.5$}
\psfrag{0.6}[c][c]{$0.6$}
\psfrag{0.7}[c][c]{$0.7$}
\psfrag{0.8}[c][c]{$0.8$}
\psfrag{1}[r][l]{$1$}
\epsfig{file=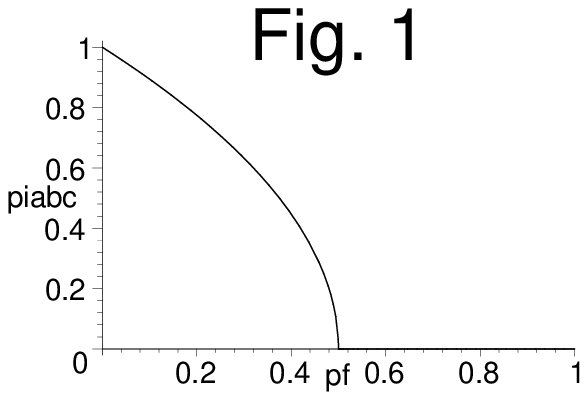, width=10cm}
\caption{}
\end{psfrags}
\end{center}
\end{figure}

\begin{figure}
\begin{center}
\begin{psfrags}
\psfrag{Fig. 2}[c][c]{}
\psfrag{pf}[t][b]{$p_f$}
\psfrag{piabc}[c][l]{$\displaystyle\frac{\cal E}{N}$}
\psfrag{0}[r][l]{$0$}
\psfrag{0.1}[c][c]{$$}
\psfrag{0.12}[c][c]{$0.12$}
\psfrag{0.14}[c][c]{$$}
\psfrag{0.16}[c][c]{$0.16$}
\psfrag{0.2}[c][c]{$0.2$}
\psfrag{0.3}[c][c]{$0.3$}
\psfrag{0.4}[c][c]{$0.4$}
\psfrag{0.5}[c][c]{$0.5$}
\psfrag{0.6}[c][c]{$0.6$}
\psfrag{0.7}[c][c]{$0.7$}
\psfrag{0.8}[c][c]{$0.8$}
\psfrag{0.08}[c][c]{$0.08$}
\psfrag{0.06}[c][c]{}
\psfrag{0.04}[c][c]{$0.04$}
\psfrag{0.02}[c][c]{}
\psfrag{\2610.08}[c][c]{$-0.08$}
\psfrag{\2610.06}[c][c]{}
\psfrag{\2610.04}[c][c]{$-0.04$}
\psfrag{\2610.02}[c][c]{}
\psfrag{1}[r][l]{$1$}
\hspace{-0.8cm}
\epsfig{file=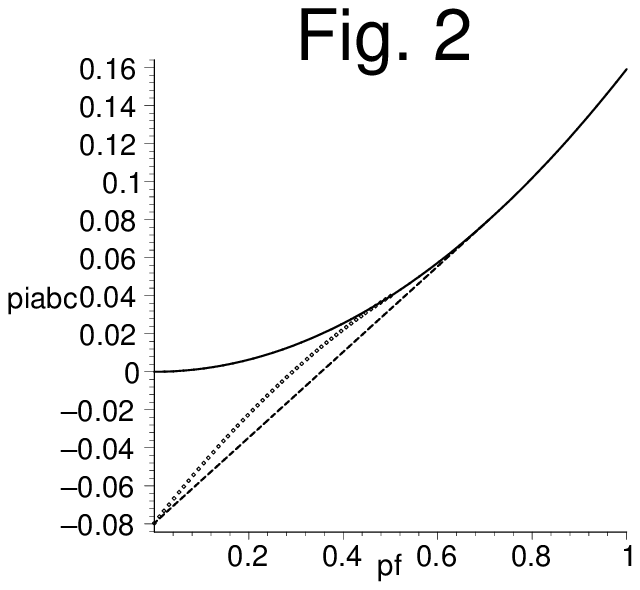, width=11cm}
\caption{}
\end{psfrags}
\end{center}
\end{figure}

\begin{figure}
\begin{center}
\begin{psfrags}
\psfrag{Fig. 3}[c][c]{}
\psfrag{pf}[t][b]{$p_f$}
\psfrag{piabc}[c][c]{$\langle \bar \psi \psi \rangle$}
\psfrag{0}[c][c]{$0$}
\psfrag{0.05}[l][r]{\hspace{-0.7cm}$0.05$}
\psfrag{0.1}[l][r]{\hspace{-0.7cm}$0.1$}
\psfrag{0.15}[l][r]{\hspace{-0.7cm}$0.15$}
\psfrag{0.2}[l][r]{\hspace{-0.7cm}$0.2$}
\psfrag{0.25}[l][r]{\hspace{-0.7cm}$0.25$}
\psfrag{0.3}[c][c]{$0.3$}
\psfrag{0.4}[c][c]{$0.4$}
\psfrag{0.5}[c][c]{$0.5$}
\psfrag{0.6}[c][c]{$0.6$}
\psfrag{0.7}[c][c]{$0.7$}
\psfrag{0.8}[c][c]{$0.8$}
\psfrag{1}[r][l]{$1$}
\epsfig{file=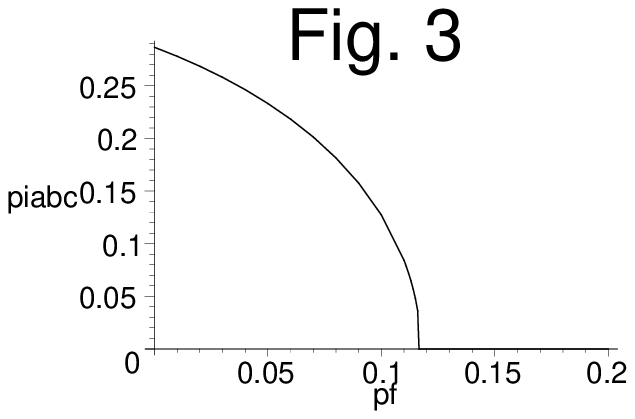, width=10cm}
\caption{}
\end{psfrags}
\end{center}
\end{figure}

\begin{figure}
\begin{center}
\begin{psfrags}
\psfrag{Fig. 4}[c][c]{}
\psfrag{xa}[tl][br]{$x-x_0$}
\psfrag{ya}[t][b]{$\langle \bar \psi \psi \rangle$}
\psfrag{za}[r][l]{$\langle \bar \psi {\rm i}\gamma^5 \psi \rangle$}
\psfrag{0}[c][c]{$0$}
\psfrag{0.05}[c][c]{$0.05$}
\psfrag{0.1}[c][c]{$0.1$}
\psfrag{0.15}[c][c]{$0.15$}
\psfrag{0.2}[c][c]{$0.2$}
\psfrag{0.25}[c][c]{$0.25$}
\psfrag{0.3}[c][c]{$0.3$}
\psfrag{0.4}[c][c]{$0.4$}
\psfrag{0.5}[c][c]{$0.5$}
\psfrag{0.6}[c][c]{$0.6$}
\psfrag{0.7}[c][c]{$0.7$}
\psfrag{0.8}[c][c]{$0.8$}
\psfrag{\2610.5}[c][c]{$-0.5$}
\psfrag{\2611}[c][c]{$-1$}
\psfrag{1}[r][l]{$1$}
\epsfig{file=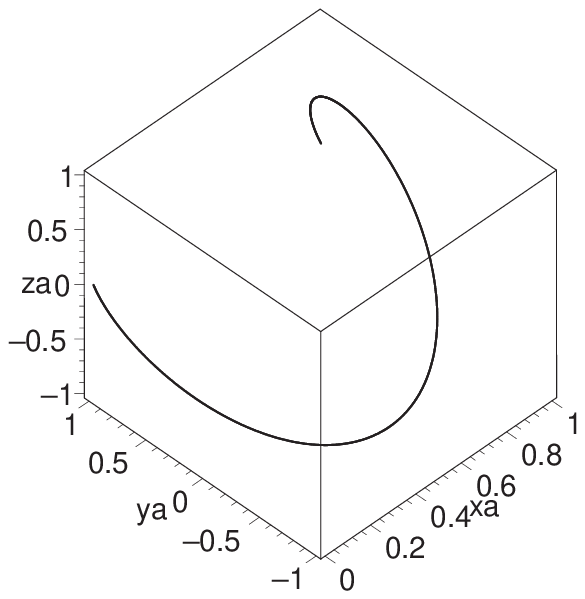, width=10cm}
\caption{}
\end{psfrags}
\end{center}
\end{figure}

\begin{figure}
\begin{center}
\begin{psfrags}
\psfrag{Fig. 5}[c][c]{}
\psfrag{xa}[tl][br]{$x-x_0$}
\psfrag{ya}[t][b]{$\langle \bar \psi \psi \rangle$}
\psfrag{za}[r][l]{$\langle \bar \psi {\rm i}\gamma^5 \psi \rangle$}
\psfrag{0}[c][c]{$0$}
\psfrag{0.05}[c][c]{$0.05$}
\psfrag{0.1}[c][c]{$0.1$}
\psfrag{0.15}[c][c]{$0.15$}
\psfrag{0.2}[c][c]{$0.2$}
\psfrag{0.25}[c][c]{$0.25$}
\psfrag{0.3}[c][c]{$0.3$}
\psfrag{0.4}[c][c]{$0.4$}
\psfrag{0.5}[c][c]{$0.5$}
\psfrag{0.6}[c][c]{$0.6$}
\psfrag{0.7}[c][c]{$0.7$}
\psfrag{0.8}[c][c]{$0.8$}
\psfrag{\2610.5}[c][c]{$-0.5$}
\psfrag{\2611}[c][c]{$-1$}
\psfrag{1}[r][l]{$1$}
\epsfig{file=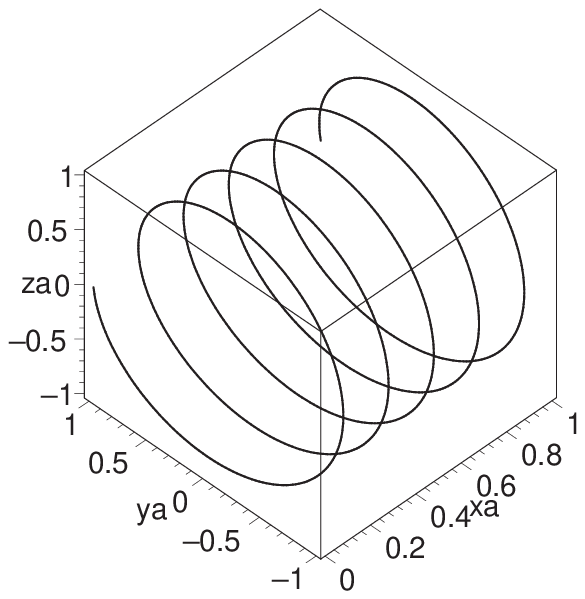, width=10cm}
\caption{}
\end{psfrags}
\end{center}
\end{figure}
\nopagebreak
\begin{figure}
\begin{center}
\begin{psfrags}
\psfrag{Fig. 6}[c][c]{}
\psfrag{pf}[t][b]{$p_f$}
\psfrag{piabc}[c][l]{$\displaystyle\frac{\cal E}{N}$}
\psfrag{0}[r][l]{$0$}
\psfrag{0.1}[c][c]{$$}
\psfrag{0.12}[c][c]{$0.12$}
\psfrag{0.14}[c][c]{$$}
\psfrag{0.16}[c][c]{$0.16$}
\psfrag{0.2}[c][c]{$0.2$}
\psfrag{0.3}[c][c]{$0.3$}
\psfrag{0.4}[c][c]{$0.4$}
\psfrag{0.5}[c][c]{$0.5$}
\psfrag{0.6}[c][c]{$0.6$}
\psfrag{0.7}[c][c]{$0.7$}
\psfrag{0.8}[c][c]{$0.8$}
\psfrag{0.08}[c][c]{$0.08$}
\psfrag{0.06}[c][c]{}
\psfrag{0.04}[c][c]{$0.04$}
\psfrag{0.02}[c][c]{}
\psfrag{\2610.08}[c][c]{$-0.08$}
\psfrag{\2610.06}[c][c]{}
\psfrag{\2610.04}[c][c]{$-0.04$}
\psfrag{\2610.02}[c][c]{}
\psfrag{1}[r][l]{$1$}
\epsfig{file=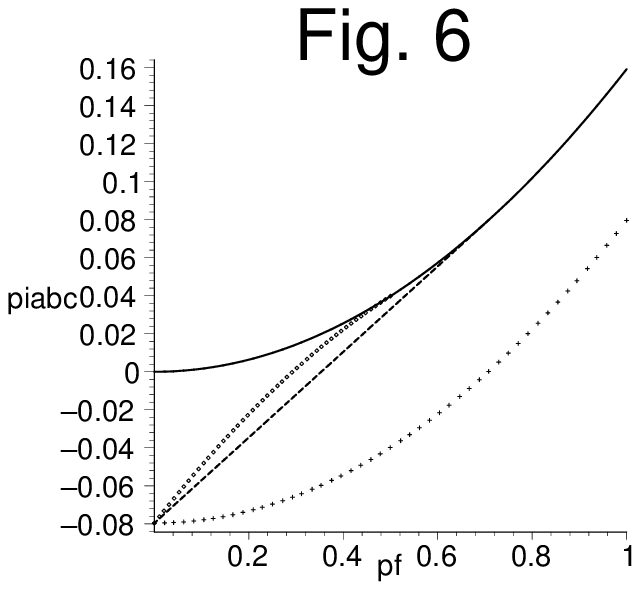, width=10cm}
\caption{}
\end{psfrags}
\end{center}
\end{figure}

\begin{figure}
\begin{center}
\begin{psfrags}
\psfrag{Fig. 7}[c][c]{}
\psfrag{pf}[t][b]{$x-x_0$}
\psfrag{piabc}[c][c]{$\displaystyle\rho_B$}
\psfrag{0}[r][l]{$0$}
\psfrag{0.1}[c][c]{$0.1$}
\psfrag{0.2}[c][c]{$0.2$}
\psfrag{0.3}[c][c]{$0.3$}
\psfrag{0.15}[c][c]{$$}
\psfrag{0.25}[c][c]{$$}
\psfrag{0.05}[c][c]{$$}
\psfrag{2}[c][c]{$2$}
\psfrag{4}[c][c]{$4$}
\psfrag{6}[c][c]{$6$}
\psfrag{8}[c][c]{8}
\psfrag{10}[c][c]{$10$}
\psfrag{0.02}[c][c]{}
\psfrag{\2610.08}[b][t]{$-0.08$}
\psfrag{\2610.06}[c][c]{}
\psfrag{\2610.04}[c][c]{$-0.04$}
\psfrag{\2610.02}[c][c]{}
\psfrag{1}[r][l]{$1$}
\epsfig{file=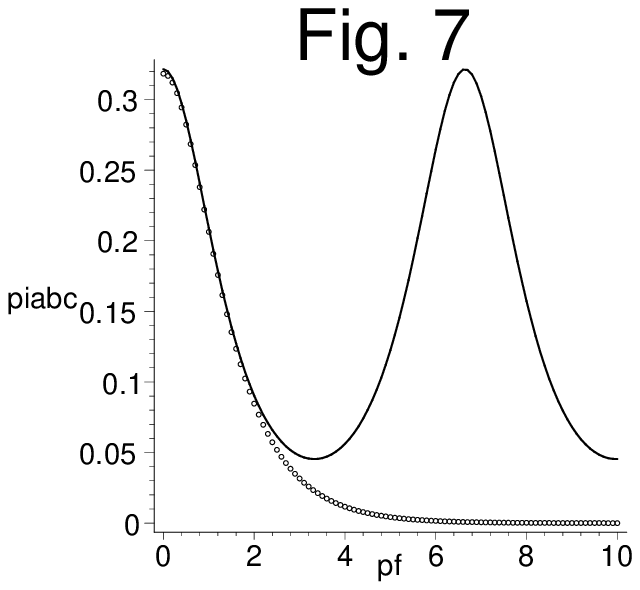, width=10cm}
\caption{}
\end{psfrags}
\end{center}
\end{figure}

\begin{figure}
\begin{center}
\begin{psfrags}
\psfrag{Fig. 8}[c][c]{}
\psfrag{pf}[t][b]{$x-x_0$}
\psfrag{piabc}[c][c]{$\langle \bar \psi \psi \rangle$}
\psfrag{0}[cc][tl]{$0$}
\psfrag{0.5}[c][c]{$0.5$}
\psfrag{2}[c][c]{$2$}
\psfrag{4}[c][c]{$4$}
\psfrag{6}[c][c]{$6$}
\psfrag{8}[c][c]{8}
\psfrag{10}[c][c]{$10$}
\psfrag{0.02}[c][c]{}
\psfrag{\2610.5}[c][c]{$-0.5$}
\psfrag{\2611}[c][c]{$-1$}
\psfrag{1}[r][l]{$1$}
\epsfig{file=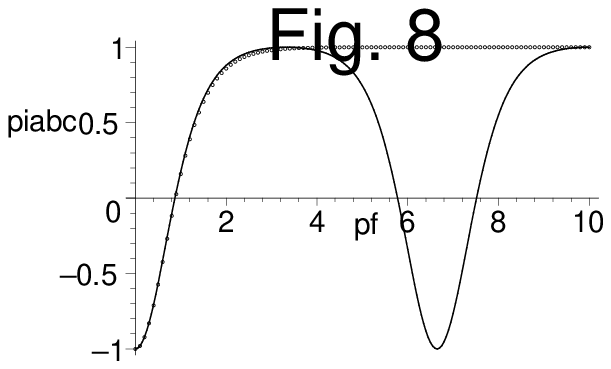, width=10cm}
\caption{}
\end{psfrags}
\end{center}
\end{figure}

\begin{figure}
\begin{center}
  \begin{psfrags}
\psfrag{Fig. 9}[c][c]{}
\psfrag{pf}[t][b]{$x-x_0$}
\psfrag{piabc}[c][l]{$\langle \bar \psi {\rm i} \gamma_5 \psi \rangle$}
\psfrag{0}[cc][tl]{$0$}
\psfrag{0.5}[c][c]{$0.5$}
\psfrag{2}[c][c]{$2$}
\psfrag{4}[c][c]{$4$}
\psfrag{6}[c][c]{$6$}
\psfrag{8}[c][c]{8}
\psfrag{10}[c][c]{$10$}
\psfrag{0.02}[c][c]{}
\psfrag{\2610.5}[c][c]{$-0.5$}
\psfrag{\2611}[c][c]{1}
\psfrag{1}[r][l]{$1$}
\epsfig{file=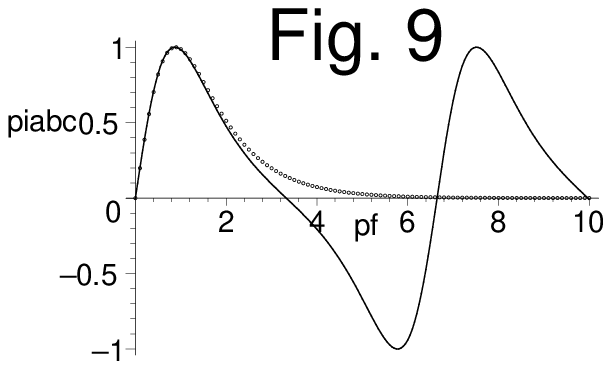, width=10cm}
\caption{}
\end{psfrags}
\end{center}
\end{figure}

\begin{figure}
\begin{center}
\begin{psfrags}
\psfrag{Fig. 10}[c][c]{}
\psfrag{xa}[tl][br]{$x-x_0$}
\psfrag{ya}[t][b]{$\langle \bar \psi \psi \rangle$}
\psfrag{za}[r][l]{$\langle \bar \psi {\rm i}\gamma^5 \psi \rangle$}
\psfrag{0}[c][c]{$0$}
\psfrag{0.05}[c][c]{$0.05$}
\psfrag{0.1}[c][c]{$0.1$}
\psfrag{0.15}[c][c]{$0.15$}
\psfrag{0.2}[c][c]{$0.2$}
\psfrag{0.25}[c][c]{$0.25$}
\psfrag{0.3}[c][c]{$0.3$}
\psfrag{0.4}[c][c]{$0.4$}
\psfrag{0.5}[c][c]{$0.5$}
\psfrag{0.6}[c][c]{$0.6$}
\psfrag{0.7}[c][c]{$0.7$}
\psfrag{0.8}[c][c]{$0.8$}
\psfrag{\2610.5}[c][c]{$-0.5$}
\psfrag{\2611}[c][c]{$-1$}
\psfrag{1}[r][l]{$1$}
\epsfig{file=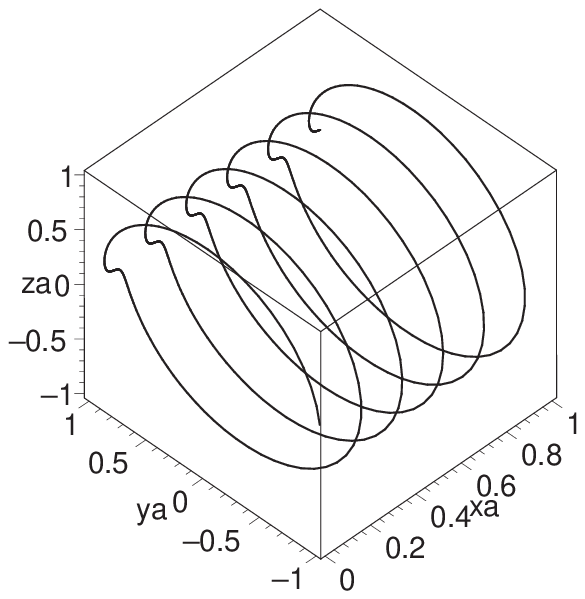, width=10cm}
\caption{}
\end{psfrags}
\end{center}
\end{figure}

\begin{figure}
\begin{center}
\begin{psfrags}
\psfrag{Fig. 11}[c][c]{}
\psfrag{xa}[tl][br]{$x-x_0$}
\psfrag{ya}[t][b]{$\displaystyle \frac{1}{p_f}$}
\psfrag{za}[r][l]{$\rho_B$}
\psfrag{0}[c][c]{$0$}
\psfrag{0.05}[c][c]{$0.05$}
\psfrag{0.1}[c][c]{$$}
\psfrag{0.15}[c][c]{$0.15$}
\psfrag{0.2}[c][c]{$0.2$}
\psfrag{0.25}[c][c]{$0.25$}
\psfrag{0.3}[c][c]{$$}
\psfrag{0.4}[c][c]{$0.4$}
\psfrag{0.5}[c][c]{$$}
\psfrag{0.6}[c][c]{$0.6$}
\psfrag{0.7}[c][c]{$$}
\psfrag{0.8}[c][c]{$0.8$}
\psfrag{0.9}[c][c]{$$}
\psfrag{4}[c][c]{$4$}
\psfrag{2}[c][c]{$2$}\psfrag{6}[c][c]{$6$}
\psfrag{8}[c][c]{$8$}
\psfrag{\2610.5}[c][c]{$-0.5$}
\psfrag{\2611}[c][c]{$-1$}
\psfrag{1}[r][l]{$1$}
\epsfig{file=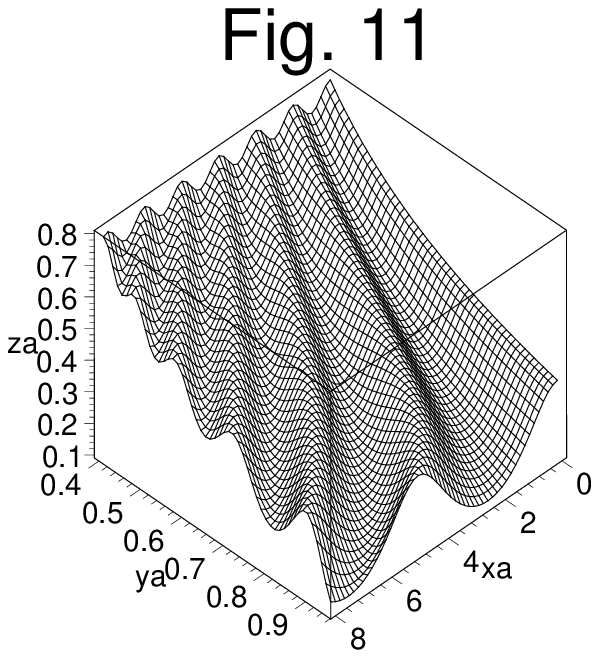, width=10cm}
\caption{}
\end{psfrags}
\end{center}
\end{figure}


\begin{thebibliography}{99}
\bibitem{Alford}
M. Alford, K. Rajagopal, and F. Wilczek, Phys. Lett. {\bf B422} (1998) 247.
\bibitem{Rapp}
R. Rapp, T. Sch\"afer, E.V. Shuryak, and M. Velkovsky, Phys. Rev. Lett.
{\bf 81} (1998) 247.
\bibitem{Nambu}
Y. Nambu and G. Jona-Lasinio, Phys. Rev. {\bf 122} (1961) 345.
\bibitem{Gross}
D.J. Gross and A. Neveu, Phys. Rev. {\bf D10} (1974) 3235.
\bibitem{tHooft74b}
G. 't~Hooft, Nucl. Phys. {\bf B75} (1974) 461.
\bibitem{tHooft74a}
G. 't~Hooft, Nucl. Phys. {\bf B72} (1974) 461.
\bibitem{Harrington}
B.J. Harrington and A. Yildiz, Phys. Rev. {\bf D11} (1975) 779.
\bibitem{Dashen}
R.F. Dashen, S. Ma, and R. Rajaraman, Phys. Rev. {\bf D11} (1975) 1499.
\bibitem{Wolff}
U. Wolff, Phys. Lett. {\bf B157} (1985) 303.
\bibitem{Treml}
T.F. Treml, Phys. Rev. {\bf D39} (1988) 679.
\bibitem{Barducci}
A. Barducci, R. Casalbuoni, M. Modugno, G. Pettini, and R. Gatto,
Phys. Rev. {\bf D51} (1995) 3042.
\bibitem{Fischler79}
W. Fischler, J. Kogut, and L.Susskind, Phys. Rev. {\bf D19} (1979) 1188.
\bibitem{Kao94}
Y-Ch. Kao and Y-W. Lee, Phys. Rev. {\bf D50} (1994) 1165.
\bibitem{Christiansen96}
H.R. Christiansen and F.A. Schaposnik, Phys. Rev. {\bf D55} (1997) 4920.
\bibitem{Deryagin92}
D.V. Deryagin, D.Yu. Grigoriev, and V.A. Rubakov,
Int. J. Mod. Phys. {\bf A7} (1992) 659.
\bibitem{Shuster99}
E. Shuster and D.T. Son, Nucl. Phys. {\bf B573}
(2000) 434.
\bibitem{Park99}
B-Y. Park, M. Rho, A. Wirzba, and I. Zahed, {\tt hep-ph/9910347}.
\bibitem{Kutschera90}
M. Kutschera, W. Broniowski, and A. Kotlorz, Nucl. Phys. {\bf A516}
(1990) 566.
\bibitem{Sadzikowski00}
M. Sadzikowski and W. Broniowski, {\tt hep-ph/0003282}.
\bibitem{McLerran}
L.D. McLerran and A. Sen, Phys. Rev. {\bf D32} (1985) 2794.
\bibitem{Li86}
Ming Li, Phys. Rev. {\bf D34} (1986) 3888.
\bibitem{Schoen}
V. Sch\"on and M. Thies, Phys. Lett. {\bf B481} (2000) 299.
\bibitem{Affleck}
I. Affleck, Nucl. Phys. {\bf B265} (1986) 448.
\bibitem{Salcedo}
L.L. Salcedo, S. Levit, and J.W. Negele, Nucl. Phys. {\bf B361} (1991) 585.
\bibitem{Lenz}
F. Lenz, M. Thies, S. Levit, and K. Yazaki, Ann. Phys. {\bf 208}
(1991) 1.
\bibitem{Coleman73}
S. Coleman, Comm. Math. Phys. {\bf 31} (1973) 259.
\bibitem{Mermin}
N.D. Mermin and H. Wagner, Phys. Rev. Lett. {\bf 17} (1966) 1133.
\bibitem{Witten78}
E. Witten, Nucl. Phys. {\bf B145} (1978) 110.
\bibitem{Berezinski}
V.L. Berezinski, Sov. Phys. JETP {\bf 32} (1971) 493.
\bibitem{Kosterlitz}
J.M. Kosterlitz and D. Thouless, J. Phys. {\bf C6} (1973) 1181.
\bibitem{Skyrme}
T.H.R. Skyrme, Proc. Roy. Soc. Lond. {\bf A260} (1961) 127.
\bibitem{Klebanov}
I. Klebanov, Nucl. Phys. {\bf B262} (1985) 133.
\bibitem{Pausch}
R. Pausch, M. Thies, and V.L. Dolman, Z. Phys. {\bf A338} (1991) 441.
\bibitem{Witten79}
E. Witten, Nucl. Phys. {\bf B160} (1979) 57.
\bibitem{Bardeen}
J. Bardeen, L.N. Cooper, and J.R. Schrieffer, Phys. Rev. Lett. {\bf 34}
(1975) 1353.
\bibitem{Chodos}
A. Chodos, R.L. Jaffe, K. Johnson, C.B. Thorn, and V.F. Weisskopf,
Phys. Rev. {\bf D9} (1974) 3471.
\bibitem{Bars}
I. Bars and M.B. Green, Phys. Rev. {\bf D17} (1978) 537.
\bibitem{Li87}
Ming Li, L. Wilets, and M.C. Birse, J. Phys. {\bf G13} (1987) 915.
\bibitem{Zhitnitsky}
A.R. Zhitnitsky, Phys. Lett. {\bf B165} (1985) 405.
\bibitem{Abdalla}
E. Abdalla and M.C.B. Abdalla, Phys. Rep. {\bf 265} (1996) 253.
\bibitem{Wu}
T.T. Wu, Phys. Lett. {\bf B71} (1977) 142.
\bibitem{Scott}
A.C.Scott, F.Y.F. Chu, and D.W. McLaughlin, Proc. IEEE {\bf 61}
(1973) 1443.
\bibitem{Oakes}
M. Gell-Mann, R.J. Oakes, and B. Renner, Phys. Rev. {\bf 175} (1968)
2195.
\bibitem{McMillan}
W.L. McMillan, Phys. Rev. {\bf B16} (1977) 4655.
\bibitem{Theodorou}
G. Theodorou and T.M. Rice, Phys. Rev. {\bf B18} (1978) 2840.
\bibitem{Takayama}
K. Takayama and M. Oka, Nucl. Phys. {\bf A551} (1993) 637.
\bibitem{Abramovitz}
{\em Handbook of Mathematical Functions}, M. Abramovitz and I.A. Stegun,
eds. (Dover, New York, 1970)

\end{thebibliography}
\end{document}